\documentclass[aps,prb,reprint]{revtex4-1}
\usepackage{graphicx}
\usepackage{amsmath}
\usepackage{amssymb}

\begin{document}

\title{Compositional bowing of band energies and their deformation potentials in
strained InGaAs ternary alloys: a first-principles study}
\author{Petr A. Khomyakov}
\email{petrk@iis.ee.ethz.ch}
\author{Mathieu Luisier}
\author{Andreas Schenk}
\affiliation{Integrated Systems Laboratory, Department of
Information Technology and Electrical Engineering, ETH Zurich,
Gloriastrasse 35, 8092 Zurich, Switzerland}

\date{\today}

\begin{abstract}
Using first-principles calculations, we show that the conduction and
valence band energies and their deformation potentials exhibit a
non-negligible compositional bowing in strained ternary
semiconductor alloys such as InGaAs. The electronic
structure of these compounds has been calculated within the
framework of local density approximation and hybrid functional
approach for large cubic supercells and special quasi-random structures,
which represent two kinds of model structures for random alloys. We
find that the predicted bowing effect for the band energy
deformation potentials is rather insensitive to the choice of the
functional and alloy structural model. The direction of bowing is
determined by In cations that give a stronger contribution to
the formation of the In$_{x}$Ga$_{1-x}$As valence band states
with $x\gtrsim 0.5$, compared to Ga cations.
\end{abstract}

\pacs{}
\keywords{}

\maketitle

III-V semiconductor compounds are considered for the integration
with Si-based microelectronics to take advantage of their high
charge carrier mobility.\cite{Wu:jms08} This has been driving basic
and applied research on various aspects of the III-V semiconductor
materials as well as on the device
design.\cite{Wu:jms08,Ionescu:nat11,Kuhn:ieee12} One field of
extensive research is to understand how the electronic properties of
the III-V compounds are altered by stress that is externally applied
to tailor characteristics of semiconductor
nanodevices.\cite{Pollak:pr68,Wu:jms08,Signorello:natcomm14}

Though III-V compounds such as InGaAs are important material systems, measuring their electronic properties is
a formidable challenge, partly due to their narrow gap. As a consequence,
rather scattered experimental data have been reported so far, especially for the $X$ and $L$ conduction band valleys, 
\cite{Vurgaftman:apr01} suggesting that reliable theoretical studies are needed.
Knowing deformation potentials of the band gap, valence band
splitting and higher energies of conduction band minima is of
particular interest for opto- and
nanoelectronics.\cite{Signorello:natcomm14,Vurgaftman:apr01} So far,
the bulk of the theoretical work has been done within the framework
of the empirical pseudopotential method combined with virtual
crystal approximation (VCA).\cite{Cohen:prb66,Nordheim:annphys31}
These methods require adjustable parameters  fitted to the existing
experimental data. Nowadays, the state-of-the-art first-principles methods
allow for large-scale atomistic, parameter-free calculations of
the properties of disordered materials\cite{Khomyakov:prl11} as well as parameter-free VCA calculations.\cite{Bellaiche:prb00} A
recent first-principles study has predicted a strong compositional
bowing of deformation potentials in group III-nitride semiconductor
compounds,\cite{Lepkowski:prb13} not supporting an assumption of a
negligible bowing.\cite{Yan:prb14} There has been
no systematic atomistic study of the compositional bowing effect in
the strained InGaAs ternary alloy, which is a channel material of
high relevance for modern types of field-effect
transistors.\cite{Wu:jms08,Ionescu:nat11,Kuhn:ieee12}

In this Letter, we have done first-principles calculations of the
band structure parameters for strained In$_{x}$Ga$_{1-x}$As
compounds as a function of alloy composition, $x$. These ternary
alloys have been modeled using 32-atom special quasi-random
structures (SQSs),\cite{Perdew:prl90,Gan:prb06,Wu:jpcc11} and
216-atom cubic supercell (SC) structures with In and Ga atoms
randomly distributed over the lattice sites.  The binary alloys
(GaAs and InAs) are first treated with a primitive 2-atom cell for
highly converged reference calculations. The supercell approach is
then adopted for correlated calculations of the physical properties
of the binary InAs and GaAs, and ternary In$_{x}$Ga$_{1-x}$As alloys
with $x=$ $0.25$, $0.50$, and $0.75$.  Using density functional
theory (DFT) approach, we have obtained the compositional dependence
for the hydrostatic deformation potentials of the band gap and
higher conduction band energies at the $L$ and $X$ valley minima as
well as two shear deformation potentials for the valence band
splitting at the $\Gamma$ point. These calculations have been done
for several random configurations of 216-atom supercell alloy
structures to compute configurational averages of the physical
quantities. We interpolate the DFT-derived compositional dependence
of a physical quantity, $Q$, as
\begin{equation}
Q(x)=x\, Q(1) + (1-x)\, Q(0) - b_{\rm w}\, x\, (1-x),\label{eq}
\end{equation}
where $b_{\rm w}$ is a bowing parameter, and $Q(0)$, $Q(x)$ and $Q(1)$ correspond to GaAs, In$_{x}$Ga$_{1-x}$As and InAs, respectively.
The bowing is assumed to be upward (downward) for $b_{\rm w}<0$ ($b_{\rm w}>0$).
Our results predict a sizable compositional bowing of band energies and their deformation
potentials ($b_{\rm w}\neq 0$), contrary to the widely used Vegard law ($b_{\rm w}=0$).\cite{Vurgaftman:apr01,Yan:prb14}
The bowing effect of the deformation potentials is attributed to a stronger contribution of the $p$-orbitals of In
cations to the formation of the alloy valence band states compared
to that of Ga for $x\sim 0.5$.

We perform the DFT calculations within the framework of the
plane-wave pseudopotential approach combined with the projector
augmented wave (PAW) formalism~\cite{Blochl:1994uk} as implemented
in the VASP code.\cite{Kresse:1996vk2,Kresse:1996vk} Local density
approximation (LDA)\cite{Perdew:prb80} and hybrid functional
approach (HSE06)\cite{Heyd:jcp06} are adopted for the DFT
calculations. The Brillouin zone (BZ) is sampled with a 12x12x12,
4x4x2, and 2x2x2 k-point grid for the 2-atom primitive cell, 32-atom
SQS, and 216-atom cubic supercell structures, respectively. The
k-meshes are chosen to include the $\Gamma$, $X$ and $L$ points in
the BZ. A kinetic energy cut-off of 400 eV is used for total energy
calculations to converge the total energy to $10^{-6}$ eV and
interatomic forces to $\sim 10^{-3}$ eV/\AA\, at least. Our
supercell calculations suggest that Vegard's law is a good
approximation for the lattice constant of the fully optimized
In$_{x}$Ga$_{1-x}$As structures, i.e., $a(x)\approx x\, a_{\rm InAs}
+ (1 - x)\, a_{\rm GaAs}$, where $a_{\rm InAs}$=6.104 \AA\, (6.032
\AA) and $a_{\rm GaAs}$=5.677 \AA\, (5.611 \AA) are obtained with
the HSE (LDA) functional. For the sake of comparison, we have also
done alloy calculations using a modern version of the virtual crystal
approximation (VCA) based on the DFT approach.\cite{Bellaiche:prb00}
This DFT-VCA scheme predicts, however, an anomalous compositional bowing of the
lattice constant and band energies in contradiction with experiment
and supercell calculations.\cite{Soekeland:prb03} Additional computational
details can be found in the supplementary material
(SM).\cite{supplmaterial}

\begin{figure}[!tpb]
\begin{center}
\includegraphics[width=1\columnwidth]{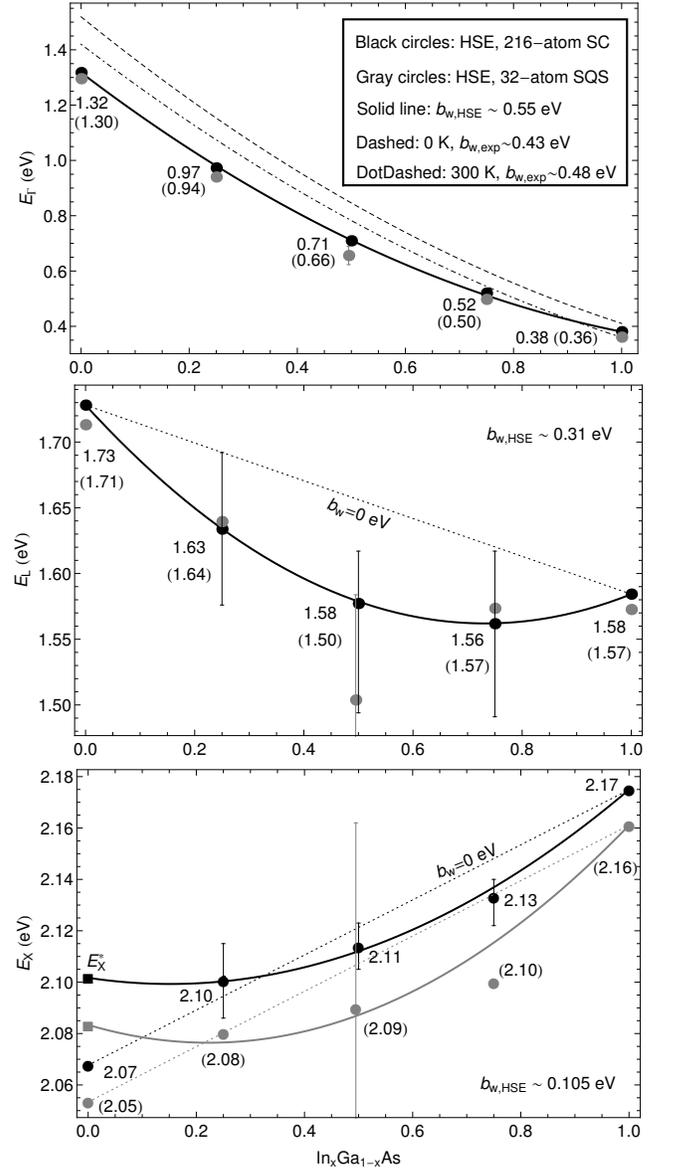}
\end{center}
\caption{Compositional dependence of the conduction band energies
$E_{\Gamma}$, $E_{L}$ and $E_{X}$ averaged over the one $\Gamma$,
four $L$, and three $X$ valley energies, respectively. The
HSE-calculated energies (black and gray filled circles) are defined
with respect to the valence band energy averaged over top three
valence bands (split by alloy disorder) at the $\Gamma$ point. The
experimental data are taken from Refs.~\onlinecite{Vurgaftman:apr01}
and \onlinecite{Goetz:jap83}. Filled squares correspond to $E_{\rm
X, GaAs}^{*}$, which is the conduction band energy of GaAs exactly
at the $X$ point of the BZ.
All the lines are given by Eq.~(\ref{eq}). The "error" bars are
given by a disorder-induced energy splitting at the effective $L$
($X$) point of the virtual fcc crystal BZ of InGaAs. {\it Note} that
these bars are not related to an error, but rather quantify a real effect
of alloy broadening. The latter are
not a measure for an uncertainty due to different atom arrangements,
which is within RT of 25 meV only.} \label{fig:energy}
\end{figure}

Using the lattice parameters, $a(x)$, we have calculated the electronic structure of the
InGaAs ternary alloys to obtain the relevant band energies at the
$\Gamma$, $L$ and $X$ valley minima as a function of alloy composition as
shown in Fig.~\ref{fig:energy}. The figure reveals a non-negligible
downward bowing for the band-gap energy ($E_{\rm g} = E_{\Gamma}$), suggesting that the Vegard
law is a rough approximation for the compositional dependence of the
InGaAs band gap. The hybrid functional calculations give a bowing
parameter of $\sim 0.55$ eV in a good agreement with that of 0.43
and 0.48 eV found in experiment at $T=0$ and 300 K,
respectively. The overall accuracy of the calculated band gap
compared to the experimental data is in the range of 4-14\%
depending on the alloy composition and temperature, and whether the
spin-orbit interaction is taken into account or not. Hereafter, we
neglect the spin-orbit interaction, assuming that it does not
significantly affect the compositional bowing of band energies and their
deformation potentials.\cite{vdWalle:prb89}  This is a plausible assumption since the
spin-orbit splitting, $\Delta_{\rm so}$, of top valence bands at the
$\Gamma$-point depends weakly on the InGaAs alloy composition as given by
$\Delta_{\rm so}(x)$, which is 0.36 and 0.38
eV for GaAs ($x=0$) and InAs ($x=1$), respectively.

To verify that the compositional bowing does not depend on a
particular choice of the alloy structural model that accounts for
configurational disorder, we explicitly show in
Fig.~\ref{fig:energy} that the 32-atom SQS and 216-atom cubic
supercell models give consistently similar values for the band gap.
The band gap calculated for the 32-atom SQSs exhibits a certain
alloy broadening for the In$_{x}$Ga$_{1-x}$As alloy with $x=0.5$
because the three top valence bands are split by alloy disorder.
Increasing the supercell size to 216 atoms reduces this effect as
seen in Fig.~\ref{fig:energy}. This suggests that some physical
quantities of the ternary alloys may suffer from a finite-size
effect as a result of treating alloy disorder with relatively-small
model structures.  It also means that actual alloy
broadening of the band gap is likely to be rather small in the limit
of an infinitely (or sufficiently) large ternary alloy structure,
but it may manifest itself in alloy nanostructures.

Figure~\ref{fig:energy} also shows how the conduction band energies of the InGaAs alloys at
the $L$ and $X$ valley minima depend on alloy composition. We find that
the conduction band energy, $E_{L}$, at the $L$-point exhibits
strong compositional bowing, suggesting that Vegard's law is not
applicable for this band parameter. The conduction band energy,
$E_{X}$, at the $X$ valley minimum has a nonlinear behavior with respect to
alloy composition as well, see Fig.~\ref{fig:energy}. The figure suggests that the alloy broadening of these energies is
more significant compared to that of the band gap.  Unfortunately, there are no reliable experimental data for the higher
conduction band energies of the In$_{x}$Ga$_{1-x}$As alloys with $x>0$ so that we provide a
parameter-free prediction of their values.

The higher energies are more prone to alloy disorder due to a
different character of the electronic states at the $L$ and $X$
valley minima compared to that of the electronic state at the
$\Gamma$ point. The latter is an isotropic, $s$-like state that
should be rather insensitive to some structural anisotropy induced
by disorder in the InGaAs alloy. The electronic states at the four
$L$ and three $X$ valleys of the BZ are anisotropic due to their
mixed $sp$-like character that makes them more sensitive to alloy
disorder. In the supercell calculations, we observe it as a four
(three)-fold degeneracy lift, i.e., energy level splitting, at the
effective $L$($X$) valley in the virtual face-centered cubic (fcc)
crystal BZ, which can be defined for the InGaAs alloys in virtual
crystal approximation. We compute the actual conduction band
energies shown in Fig.~\ref{fig:energy} as an average of the
corresponding four (three) energies at the effective $L$ ($X$)
valley. The alloy broadening of the band energies is proportional to
the disorder-induced energy level splitting, see
Fig.~\ref{fig:energy}. Though this splitting can be sizeable, the
averaged conduction band energies are insensitive to configurational
disorder with some exception for $x\sim 0.5$. That is demonstrated
by using two different structural models of random alloys (the SQS
and cubic supercells) that give similar averaged energies as shown
in Fig.~\ref{fig:energy}. Thus, we provide a reliable prediction for
the average band structure of InGaAs even though alloy broadening
might be overestimated because of the finite-size effect.

\begin{figure}[!tpb]
\begin{center}
\includegraphics[width=1\columnwidth]{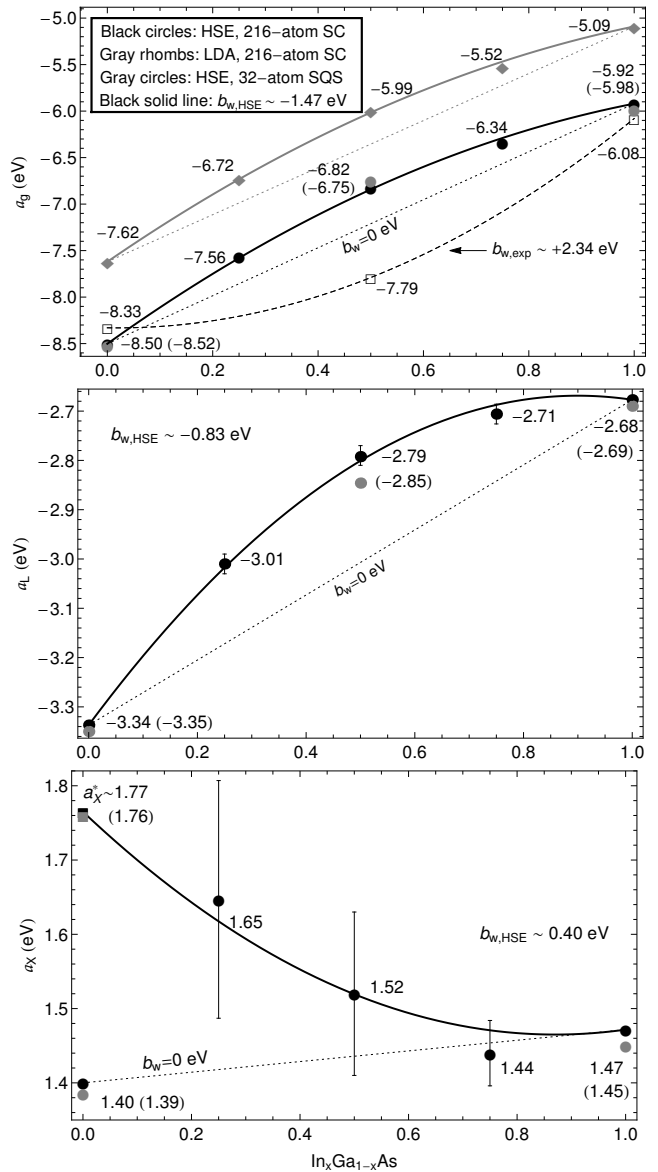}
\end{center}
\caption{Compositional dependence of the hydrostatic deformation
potentials ($a_{\Gamma}$, $a_{L}$ and $a_{X}$) of the conduction
band energies that are defined as $a_{\rm C}=[a_{\rm C}(\epsilon=0+)
+ a_{\rm C}(\epsilon=0-)]/2$ with ${\rm C}=\Gamma$, $L$ and $X$,
where $a_{\rm C}(\epsilon=0+)$ and $a_{\rm C}(\epsilon=0-)$ are
calculated with the HSE (filled circles and squares) and LDA (filled
gray rhombs) functionals for the tensile and compressive strain
conditions, respectively. Filled squares correspond to the
deformation potential, $a_{\rm X}^{*}$, for the conduction band
energy of GaAs exactly at the $X$ point of the BZ. The "error" bars
correspond to alloy disorder-induced anisotropy quantified as $\vert
a_{\rm C}(\epsilon=0+) - a_{\rm C}(\epsilon=0-)\vert$. The open
squares are related to the experimental data.\cite{Vurgaftman:apr01} All
the lines are given by Eq.~(\ref{eq}).} \label{fig:defpot}
\end{figure}
We now describe how the conduction band energies are affected by
hydrostatic strain. The corresponding deformation potential, $a_{\rm
C}$, is given by $E_{\rm C}(\epsilon)=E_{\rm C}(0)+3\, a_{\rm
C}\,\epsilon$ with $C=\Gamma$, $L$ and $X$, where
$\epsilon=\epsilon_{xx}=\epsilon_{yy}=\epsilon_{zz}\ne 0$ are the
only non-zero elements of the strain
tensor.\cite{vdWalle:prb89,supplmaterial} A larger absolute value of
the deformation potential means a stronger strain influence on the
alloy band structure, and the sign of $a_{\rm C}$ suggests a
qualitative behavior of the band energies with respect to the
compressive ($\epsilon<0$) and tensile ($\epsilon>0$) strain
conditions. Using the LDA and HSE functionals, we have calculated
the band-gap hydrostatic deformation potential, $a_{\Gamma}$, for
the 216-atom cubic supercell structures of the In$_{x}$Ga$_{1-x}$As
ternary alloys as a function of alloy composition as given in
Fig.~\ref{fig:defpot}. The LDA (HSE) calculations agree (very well)
with the experimental data for the GaAs and InAs binary alloys. For
the InGaAs ternary alloys, a non-negligible {\it upward} bowing is
predicted for $a_{\Gamma}$, see Fig.~\ref{fig:defpot}. The 32-atom
SQS model shows a similar upward behavior for the compositional
bowing of $a_{\Gamma}$, proving that we have done configurational
averaging of this physical parameter properly.

Having a good agreement with experiment for the two binary alloys
(GaAs and InAs) suggests that the HSE calculations should
allow for reliable prediction of hydrostatic deformation potentials
for the corresponding ternary alloys. Though three experimental data
points for the In$_{x}$Ga$_{1-x}$As alloys with $x=0.0$, $0.5$, and
$1.0$ (Fig.~\ref{fig:defpot}) seem to suggest a {\it downward}
compositional bowing of the band-gap hydrostatic deformation
potential, we would like to point out that these data are taken from
three uncorrelated sources.\cite{Vurgaftman:apr01}
Another explanation is that determining $a_{\rm g}$ from row
experimental data requires separating the hydrostatic and shear
strain effects on the band gap. This procedure relies on the
knowledge of elastic constants for InGaAs, which are usually
estimated from Vegard's law. However, we have found that the elastic
constant compositional dependence of the InGaAs compounds deviates
from the Vegard law as discussed in SM,\cite{supplmaterial} and this
might be a reason for the discrepancy.

Besides the band-gap energy of the InGaAs compounds, the hydrostatic
strain affects the higher energies at the $L$ and $X$ valley minima.
The corresponding deformation potentials are given as a function of
alloy composition in Fig.~\ref{fig:defpot}. One can clearly see that
there exists a strong upward (downward) bowing of the hydrostatic
deformation potential, $a_{L}$ ($a_{X}$), for the conduction band
energy exactly at the $L$ ($X$) point. For the $X$ valley minimum,
$a_{X}(x)$ exhibits a non-monotonic behavior, reaching its maximum
at $x<0.25$. We find that the deformation potential, $a_{L}$,
exhibits a modest alloy-induced anisotropy (proportional to alloy
broadening) given with "error" bars in Fig.~\ref{fig:defpot}. That
is confirmed by the HSE calculations done for the 32-atom SQS and
216-atom cubic supercell structures that represent two different
models for configurational disorder in the InGaAs ternary alloys. On
the contrary, the hydrostatic deformation potential, $a_{X}$,
suffers from a significantly larger alloy broadening than $a_{L}$.
That is mainly due to strain-induced chemical hybridization of the
electronic states at the effective $X$-point of the supercell
(unfolded) BZ.

\begin{figure}[!tpb]
\begin{center}
\includegraphics[width=1\columnwidth]{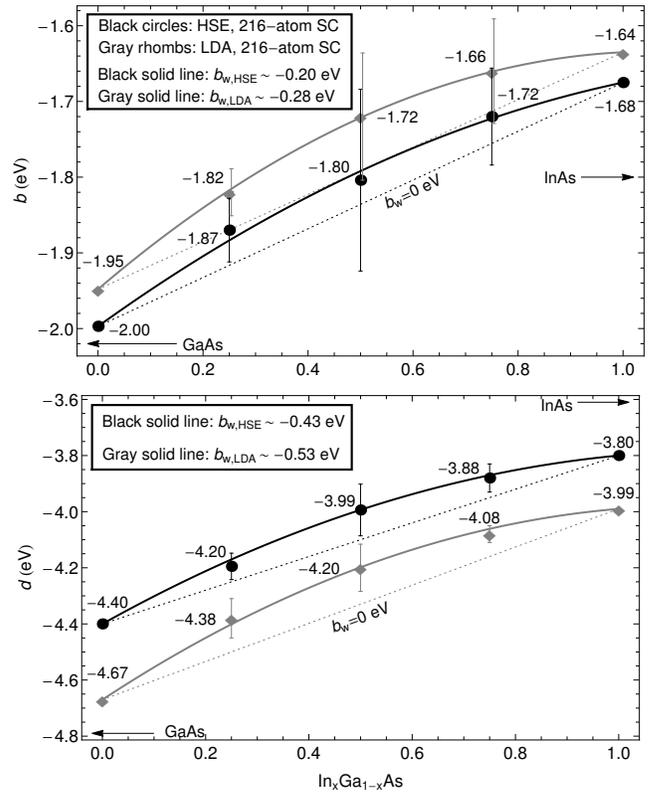}
\end{center}
\caption{Compositional dependence of the shear deformation
potentials, $D=b$ or $d$, defined as $D=[D(\epsilon=0+) +
D(\epsilon=0-)]/2$, where $D(\epsilon=0+)$ and $D(\epsilon=0-)$ are
calculated for the tensile and compressive strain conditions, using
the LDA (gray filled rhombs) and HSE (black filled circles)
functionals.  All the lines are given by Eq.~(\ref{eq}). The
recommended values from Ref.~\onlinecite{Vurgaftman:apr01} are given
with arrows. The "error" bars are related to disorder-induced
anisotropy, $\vert D(\epsilon=0+) - D(\epsilon=0-)\vert$.}
\label{fig:shear_defpot}
\end{figure}
The band structure of the In$_{x}$Ga$_{1-x}$As ternary alloy can
also be modified by two shear deformations (I) $\epsilon =
\epsilon_{xx} - \epsilon_{yy}$, $\epsilon_{yy}=-\epsilon_{xx}\ne 0$
and $\epsilon_{zz}=0$, and (II) $\epsilon =
\epsilon_{xy}=\epsilon_{yx}\ne 0$.\cite{vdWalle:prb89,Yan:prb14} One
of three top valence bands at the $\Gamma$ point is not altered by
shear strain whereas two other valence bands are split symmetrically
by the strain. The corresponding splitting energy is given as
$\Delta E_{\rm v}=D\,\vert\epsilon\vert$ in the linear
approximation, which we find to be valid up to $\epsilon\sim 0.01$
at least.  We define two shear deformation potentials as $D_{\rm I}
=3b$ and $D_{\rm II} =2\times 3^{1/2}d$, where $b$ and $d$
correspond to the shear strain I and II according to $k\cdot p$
model conventions.\cite{Picus:jetp62,Pollak:pr68} The microscopic
parameters obtained from our DFT calculations can then be directly
employed for this phenomenological model. The shear strain affects
the band gap by decreasing it to $E_{g}(0)-\vert \Delta E_{\rm v}
\vert/2$ unlike the hydrostatic strain that can enlarge or narrow
the gap, depending on the sign of $\epsilon$. The electronic states
at the $\Gamma$ ($X$, $L$) valley are unaffected (somewhat affected)
by shear strain, see SM.\cite{supplmaterial}

Figure~\ref{fig:shear_defpot} shows how the shear deformation
potentials ($b$ and $d$) depend on alloy composition. One can see
that the difference between the LDA and HSE-calculated $b$ ($d$)
deformation potentials is of $\sim$3\% (5\%) only. This is likely
due the fact that the valence bands are usually well described by
the LDA approximation.\cite{vdWalle:prb89} In
Fig.~\ref{fig:shear_defpot}, we give some values of $b$ ($d$)
recommended in Ref.~\onlinecite{Vurgaftman:apr01} for the binary
alloys. Our HSE calculations agree well with these recommended data,
predicting a somewhat higher (lower) shear deformation potential for
InAs (GaAs) by $\sim$10\% (0\%) and $\sim$5\% (8\%) for the $b$ and
$d$ deformation potentials, respectively. However, we would like to
emphasize that there hardly exist reliable experimental data for a
proper comparison. The present first-principles study does not rely
on any adjustable parameters. It can serve as a reference for future
experiments, and can be exploited to test empirical band structure
calculations.

Using the first-principles calculations, we can now clarify whether
the compositional bowing of shear deformation potentials is strong
or not. Figure~\ref{fig:shear_defpot} clearly shows a nonlinear
compositional dependence of the $b$ and $d$ parameters. This bowing
is not as strong as that predicted for group-III nitride alloys in
Ref.~\onlinecite{Lepkowski:prb13}, but it reveals the apparent
limitation of Vegard's law, which is often assumed for the band
energy deformation potentials of ternary
alloys.\cite{Vurgaftman:apr01,Yan:prb14} We find that the
compositional bowing of the deformation potentials is upward in the
In$_{x}$Ga$_{1-x}$As alloys, meaning that the strain dependence for
the valence band splitting of ternary alloys with $x\sim 0.5$ is
more similar to that of InAs (not GaAs). A plausible explanation for
this behavior is that the top valence bands, which show a $p$-like
character, are preferably formed by the $p$-orbitals of In (not Ga)
cation atoms in the In$_{x}$Ga$_{1-x}$As alloys with $x\gtrsim 0.5$.
This is confirmed by comparing the relative contribution of In and
Ga atomic orbitals to the density of states for the top valence
bands. Finally, we notice that there exists some broadening of the
$b$ and $d$ parameters because of the disorder-induced anisotropy
with respect to the strain parameter sign, see
Fig.~\ref{fig:shear_defpot}. This broadening of $d$ ($b$) is,
however, weaker than (comparable to) the compositional bowing of $d$
($b$). The anisotropy becomes unphysically large for the shear
deformation potentials obtained with the 32-atom SQS structures (not
reported in Fig.~\ref{fig:shear_defpot}) because of a finite-size
effect.

 In conclusion, we studied the structural, electronic and
mechanical properties of strained In$_{x}$Ga$_{1-x}$As ternary
alloys in a systematic manner, using a first-principles approach for
various alloy compositions and strain configurations. The
compositional and strain dependencies of the band structure
parameters as well as other relevant material parameters (tabulated
in SM, see Ref.~\onlinecite{supplmaterial}) were obtained for these
semiconductor compounds, allowing for device simulations within the
framework of the tight-binding and drift-diffusion models. We found
that the band energies and their deformation potentials can exhibit
a significant compositional bowing. Using the energy and
atom-resolved density of states, the role of In cation atoms was
revealed in relation to the bowing direction of the InGaAs
deformation potentials. That suggested a simple way to understand
the bowing effect in semiconductor alloys in a qualitative manner.

\begin{acknowledgments} This research is funded by the EU Commission (FP7 project: DEEPEN).
The computer simulations are done at the Swiss National
Supercomputer Center (project: s579).
\end{acknowledgments}

\bibliography{InGaAs_Letter}

\begin{center}
SUPPLEMENTARY MATERIAL
\end{center}

\begin{center}
{\it Hybrid functional calculations}
\end{center}
To calculate the electronic, structural and mechanical properties of
the strained InGaAs compounds, we employ the conventional HSE06
hybrid functional functional with 25\% of the short-range Fock
exchange and the range-separation parameter of $\mu=0.207$
\AA$^{-1}$ for all alloy compositions. However, ion relaxation is
first done using the semi-local GGA-PBE functional to have a good
initial guess for atomic positions in alloy structures.
Subsequently, we perform total energy calculations with the HSE06
functional to achieve a better convergence for the atomic and
electronic structure of alloys. Even though this second ion
optimization procedure usually affects the HSE-calculated electronic
structure marginally, it is particularly important for an accurate
calculation of the strain II deformation potential, $d$. The shear
strain II induces an internal strain in the InGaAs alloy structures,
giving rise to a piezoelectric effect. Accounting for this effect
properly requires calculating the interatomic forces and band
energies in a consistent manner, i.e., with the same functional for
all the physical quantities.

For a 216-atom supercell, we have found that the band structure
energies are sufficiently accurate if a lower kinetic energy cut-off
of 300 eV and a $\Gamma$-point for the short-range Fock exchange
part of a hybrid functional are adopted. That allows reducing the
computational burden of hybrid functional calculations without
compromising the required accuracy. The induced error (upward shift)
is of $\sim$20 meV, which is given by the difference between the
32-atom SQS (2-atom primitive cell) and 216-atom
supercell-calculated band parameters for the binary alloys as seen
in Fig.~1 in the main text and Table~SI in supplementary material.
This shift is within an acceptable error bar since it is much
smaller than the difference between the calculated and experimental
values of the band energies.

\begin{table*}[h]
\centering \caption{Material parameters of the InGaAs compounds
calculated using the HSE06 functional. The conduction band energies
of the binary alloys (GaAs and InAs), $E_{\Gamma}$, $E_{L}$,
$E_{X}$, and $E_{X}^{*}$ (in eV), as well as the hydrostatic and
shear deformation potentials, $a_{\Gamma}$, $a_{L}$, $a_{X}$,
$a_{X}^{*}$, $b$ and $d$ (in eV), are calculated for a 2-atom
primitive cell without the spin-orbit coupling included. The
spin-orbit splitting, $\Delta_{\rm so}$, given in the table allows
accounting for the spin-orbit interaction effect on the band
energies within a $k\cdot p$ model. In the parenthesis, we report
the conduction band energies obtained with the spin-orbit coupling
included; the experimental lattice constants have been adopted in
this case.  The lattice parameters, $a$, and elastic constants,
$C_{11}$, $C_{12}$ and $C_{44}$, are given in units of \AA\, and
GPa, respectively. Any material parameter, $Q$, for a given ternary
alloy composition, $x$, can be obtained using the interpolation
formula Eq.~(1) in the main text, where the bowing parameters,
$b_{\rm w}$, are obtained using 216-atom supercell model structures
for the InGaAs ternary alloys. The compositional dependence of the
band energy, $E_{X}$, at the $X$ valley minimum can be approximated
as $E_{X}(x)=x\, E_{X,{\rm InAs}} + (1-x)\, E_{X,{\rm GaAs}}^{*} -
b_{\rm w}\, x (1-x)$ for $x>0.25$. The same approximation can be
adopted for $a_{X}$.  As an example, we give the parameters for the
most relevant alloy, In$_{0.53}$Ga$_{0.47}$As.  The experimental
data are given at room temperature, Vurgaftmana {\it et al} [Appl.
Phys. Rev. {\bf 89}, 5815 (2001)] and K.-H. Goetz {\it et al} [J.
Appl. Phys. {\bf 54}, 4543 (1983)].} \hfill{}
\begin{tabular*}{\textwidth}{l @{\extracolsep{\fill}}ccccccccccccccc}
\hline\hline
& $E_{\Gamma}$ & $E_{L}$ & $E_{X}$ & $E_{X}^{*}$ & $\Delta_{\rm so}$ & $a_{\Gamma}$ & $a_{L}$ & $a_{X}$ & $a_{X}^{*}$ & $b$   & $d$   & $C_{11}$ & $C_{12}$  & $C_{44}$  & $a$ \\
\hline
 GaAs                     & 1.29  &  1.72   &  2.05   &  2.09   & 0.36   & -8.54  & -3.35   & 1.39 & 1.76    & -2.01 & -4.33 & 117.5    & 50.0      & 60.3      & 5.677 \\
                          & (1.32) &  (1.65)  &  (1.91)   &   (1.95)   &  (0.36)  &   &   &  &     &  &  &    &       &      &  \\
 Exp                      & 1.42  &  1.81   &  1.97   & -       & 0.34   & -8.33  & -       & -    & -       & -2.0  & -4.8  & 119.0    & 53.4      & 59.6      & 5.653 \\
\hline
 In$_{0.53}$Ga$_{0.47}$As & 0.66  &  1.57   &  2.11   & 2.11    & 0.37   & -6.81  & -2.79   & 1.50 & 1.50    & -1.79 & -3.90 & 96.5     & 46.9      & 45.2      & 5.891 \\
                          & (0.69)  &  (1.51)   &  (1.95)   & (1.95)  &  (0.37)   &   &    &  &     &  &  &      &       &       &  \\
 Exp                      & 0.75  & -       & -       & -       & -      & -7.79  & -       & -    & -       & -     & -     & -        & -         & -         & 5.868 \\
\hline
 InAs                     & 0.36  &  1.58   &  2.17   & 2.17    & 0.38   & -5.97  & -2.69   & 1.45 & 1.45    & -1.69 & -3.72 & 85.2     & 45.2      & 40.3      & 6.104 \\
                          & (0.39) & (1.53)  & (2.00)  & (2.00) & (0.38) &   &    &  &     &  &  &      &       &      &  \\
 Exp                      & 0.35  & -       & -       & -       & 0.39   & -6.08  & -       & -    & -       & -1.8  & -3.6  & 83.4     & 45.4      & 39.5      & 6.058 \\
\hline
$b_{\rm w}$               & 0.55  &  0.31   & -       & 0.105   & 0      & -1.47  & -0.83   & -    & 0.40    & -0.20 & -0.43 & 15.5     & 2.2       & 17.9      & $\sim$0 \\
Exp                       & 0.48  & -       & -       & -       & -      &  2.34  & -       & -    & -       & -     & -     & -        & -         & -         & $\sim$0 \\
\hline\hline
  \end{tabular*}
\hfill{}
\label{tab}
\end{table*}

\begin{table*}[h]
\centering \caption{Hydrostatic and shear deformation potentials of the InGaAs compounds
calculated using the HSE06 functional for the conduction band
valleys, $\Xi_{d,\Gamma}-a_{\rm V}$, $\Xi_{d,X}-a_{\rm V}$,
$\Xi_{u,X}$, $\Xi_{d,L}-a_{\rm V}$ and $\Xi_{u,L}$ in units of eV,
where $a_{\rm V}$ is the hydrostatic deformation potential for the
average of three top valence bands at the $\Gamma$ point. We note that $\Xi_{u,\Gamma}=0$. The
experimental data are taken from {\it Landolt-B\"ornstein, Numerical
Data and Functional Relationships in Science and Technology}
(Springer, New York, 1982), Group III, Vol. 17a-b. In the
parenthesis and last column, we give the theoretical values of
deformation potentials, $\Xi_{u}^{\rm LDA}$ and $a_{\rm V}^{\rm
LDA}$, which were obtained using the DFT-LDA calculations for the
binary alloys (GaAs and InAs) by Chris Van de Walle [Phys. Rev. B
{\bf 39}, 1871 (1989)], and Vegard's law is used to estimate the
value of $a_{\rm V}$ for the In$_{0.53}$Ga$_{0.47}$As alloy. The
compositional dependence of the deformation potential, $\Xi_{u,X}$,
can be approximated as $\Xi_{u,X}(x)=x\, \Xi_{u,X,{\rm InAs}} +
(1-x)\, \Xi_{u,X,{\rm GaAs}}^{*} - b_{\rm w}\, x (1-x)$ for
$x>0.25$. } \hfill{}
 \begin{tabular*}{\textwidth}{l @{\extracolsep{\fill}}ccccccccc}
 \hline\hline
& $\Xi_{d,\Gamma}-a_{\rm V}$ & $\Xi_{d,X}-a_{\rm V}$ & $\Xi_{u,X}$ & $\Xi_{d,X}^{*}-a_{\rm V}$ & $\Xi_{u,X}^{*}$ & $\Xi_{d,L}-a_{\rm V}$ & $\Xi_{u,L}$ & $\Xi_{u,L}^{\rm exp}$ & $a_{\rm V}^{\rm LDA}$ \\
 \hline
  GaAs                     & -8.54 & -1.35 & 8.22 (8.61)  & -0.34 & 6.30    & -8.38  & 15.1 (14.26)  & 19.6 & 1.16\\
  In$_{0.53}$Ga$_{0.47}$As & -6.81 & -0.22 & 5.15         & -0.22 & 5.15    & -7.09  & 12.9          & - & 1.08 \\
  InAs                     & -5.97 & -0.16 & 4.83 (4.50)  & -0.16 &  4.83   & -6.86  & 12.5 (11.35)  & - & 1.00 \\
 \hline
 $b_{\rm w}$               & -1.47 & -     & -            & -0.10 & 1.50    &  -1.96 & 3.38          & - & - \\
 \hline
 \hline
   \end{tabular*}
 \hfill{}
 \label{tab2}
 \end{table*}

\begin{center}
{\it Lattice parameter definition for ternary alloys}
\end{center}
 To understand the structural properties of
the InGaAs compounds, we have adopted two structural models (the
32-atom SQS and 216-atom cubic supercells) as discussed in the
manuscript. In both cases, an effective 2-atom cubic cell is assumed
for calculating a lattice parameter, $a$, of ternary alloys, which
is derived from a simple relation between the supercell volume
($\Omega$) and lattice constant ($a$), $\Omega=N_{\rm u}\,a^{3}$,
where $N_{\rm u}$ is the number of 2-atom cells constituting the SQS
or cubic supercell structures. To calculate the lattice parameter, a
full structural optimization has been done for the ternary alloy
model structures. We have found essentially no compositional bowing
of the lattice constant for the fully optimized InGaAs structures,
i.e., $b_{\rm w}\sim 0$. Doing ion relaxation is crucial to account
for the effect of alloy disorder in ternary alloys properly.
Otherwise, it would lead to the wrong prediction of an upward
($b_{\rm w}<0$) compositional bowing for the lattice constant of the
InGaAs ternary alloys in spirit of the VCA that does not allow any
ion relaxation. Note that the HSE-calculated lattice constant for
GaAs (InAs) is just 0.5\% (0.8\%) larger than the corresponding
experimental parameter as seen in Table~SI. Thus, the hybrid
functional approach can be expected to give the lattice parameters
of the InGaAs alloys with a similar accuracy.

\begin{center}
{\it Compositional dependence of elastic constants}
\end{center}
Another explanation for the apparent discrepancy between the
DFT-calculated and experimental values of the band-gap deformation
potential is that the measurement for the In$_{x}$Ga$_{1-x}$As alloy
with $x=0.53$ was done under general strain conditions. It means
that it requires separating the hydrostatic and shear strain effects
on the band-gap energy to determine the actual hydrostatic
deformation potential. This separation procedure relies on the
knowledge of other alloy material parameters such as elastic
constants, which are usually assumed to be given by Vegard's law.
However, we have found that the elastic constants of the InGaAs
compounds exhibit some downward compositional bowing that might need
to be taken into consideration for more accurate interpretation of
the experimental data. In Table~SI, we report the elastic constants
of the binary alloys (GaAs and InAs) as well as the corresponding
compositional bowing parameters for the InGaAs ternary alloy
calculated with the HSE functional. The table shows that the elastic
constants of the binary alloys agree very well (within $\sim$1-2\%)
with the existing experimental data. A somewhat larger deviation of
$\sim$6\% is found for $C_{12, {\rm GaAs}}$.

\begin{center}
{\it Conduction band deformation potentials}
\end{center}
We would like to notice that the conduction band hydrostatic
deformation potentials, $a_{\Gamma}$, $a_{L}$ and $a_{X}$, in
Table~SI can be expressed as $a_{\Gamma}=\Xi_{d,{\Gamma}}-a_{\rm V}$
and $a_{C}=\Xi_{d,{\rm C}}+\Xi_{u,{\rm C}}/3-a_{\rm V}$, where $C=L$
or $X$, and $\Xi_{d,{\Gamma}}$, $\Xi_{d,{\rm C}}$ and $\Xi_{u,{\rm
C}}$ are the deformation potentials as defined by Herring and Vogt
[Phys. Rev. {\bf 101}, 944 (1956)], and $a_{\rm V}$ is the
hydrostatic deformation potential for the average of three top
valence bands at the $\Gamma$ point. The electronic states at the
$\Gamma$ and $L$ ($\Gamma$ and $X$) valleys of the face-centered
cubic BZ are not affected by the shear strain I (II), being
protected by symmetry, i.e., $\Xi_{u,\Gamma}=0$. The three $X$ (four
$L$) valleys are split by the shear strain I (II) into three
singlets (doublets) with a shift of $\Delta E_{X}=0$ and $\Delta
E_{X}=\pm\Xi_{u,X}\vert\epsilon \vert$ ($\Delta E_{L}=\pm
2\,\Xi_{u,L}\,\vert\epsilon \vert/3$) with respect to the averaged
conduction band energy. The conduction band deformation potentials
are tabulated in Table~SII.

\begin{center}
{\it VCA vs. supercell calculations of ternary alloys}
\end{center}
For the sake of comparison, we have done explicit virtual crystal
approximation (VCA) calculations based on density functional theory
(DFT), using a version of DFT-VCA proposed by Bellaiche and
Vanderbilt [Phys. Rev. B {\bf 61}, 7877 (2000)]. In Fig.~S1, the obtained
results show that this DFT-VCA scheme predicts an anomalous downward
compositional bowing of the lattice constant for InGaAs ternary
alloys in consistency with findings of S\"okeland and co-workers
[Phys. Rev. B {\bf 68}, 075203 (2003)] for InGaN ternary alloys. It
means that the DFT-VCA fails to reproduce the linear compositional
dependence (no bowing) of the lattice parameter of InGaAs suggested by
experiment and our supercell calculations. Using the DFT-VCA also
results in an unphysical downward compositional bowing of the band
gap and higher conduction band energies as seen in Fig.~S1. For
example, the compositional dependence of the band gap energy has a
significant contribution of the third-order correction ($x^{3}$) in contradiction with the
supercell calculations and experimental data. If we impose no bowing
for the lattice parameter in the DFT-VCA calculations (strained VCA), the compositional bowing of the band gap
increases significantly, showing even stronger deviation from
the experimental findings and supercell calculations, see Fig.~S1. S\"okeland and
co-workers have proposed that these failures of DFT-VCA are due to
rather different lattice constants of alloy components (InN and GaN
in their case, and InAs and GaAs in our case) as well as a very
different localization of the In and Ga electronic states that are
to be mixed in the VCA calculations.

\begin{figure}[!tpb]
\begin{center}
\includegraphics[width=1\columnwidth]{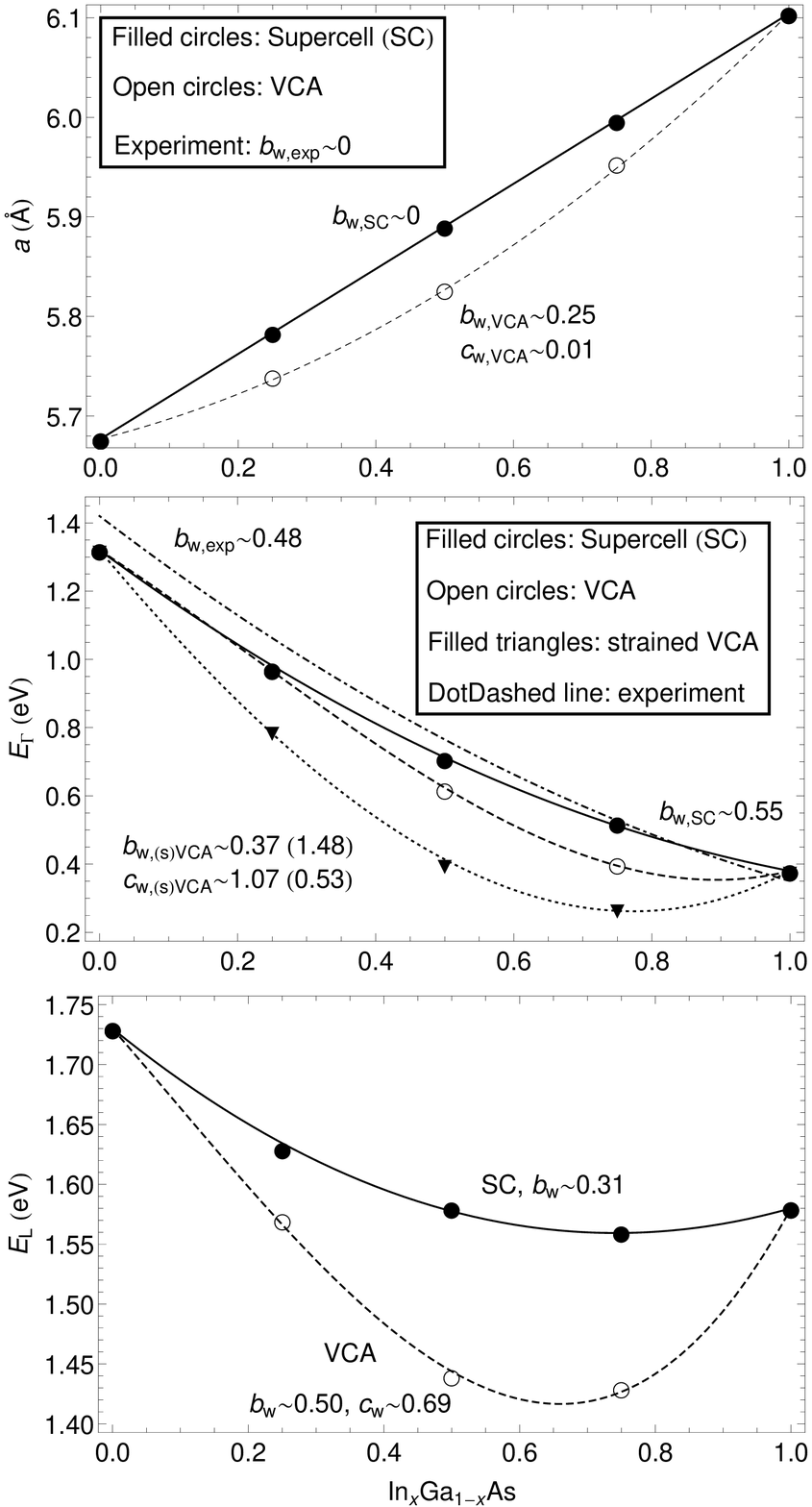}
\end{center}
\caption{Compositional dependence of the lattice constant, $a$ (top), band gap energy (middle), and $L$-valley conduction band energy (bottom) calculated using the HSE functional for two alloy models based on (i) 216-atom supercell (SC), and (ii) 2-atom virtual crystal approximation (VCA) by Bellaiche and
Vanderbilt. Filled (open) black circles correspond to supercell (VCA) calculations. Filled black triangles correspond to strained VCA calculations where the supercell-calculated lattice constant is adopted, i.e., assuming no bowing for the lattice constant. The solid and dotdashed black lines are given by the formula Eq.~(1), and the dashed and dotted black lines are given by a modified formula Eq.~(1) where a third-order term, $-c_{\rm w}\, x^{2} (1-x)$, is added for having a better fit of VCA data points. In parenthesis, we give bowing parameters for strained VCA.} \label{fig:vca}
\end{figure}

\end{document}